# Multiple Bifurcations in the Periodic Orbit around Eros

Yanshuo Ni[1], Yu Jiang[1, 2], Hexi Baoyin[1]

1. School of Aerospace Engineering, Tsinghua University, Beijing 100084, China
2. State Key Laboratory of Astronautic Dynamics, Xi'an Satellite Control Center, Xi'an 710043, China

Y. Jiang (✉) e-mail: jiangyu_xian_china@163.com (corresponding author)

H. Baoyin (✉) e-mail:baoyin@tsinghua.edu.cn (corresponding author)

**Abstract**. We investigate the multiple bifurcations in periodic orbit families in the potential field of a highly irregular-shaped celestial body. Topological cases of periodic orbits and four kinds of basic bifurcations in periodic orbit families are studied. Multiple bifurcations in periodic orbit families consist of four kinds of basic bifurcations. We found both binary period-doubling bifurcations and binary tangent bifurcations in periodic orbit families around asteroid 433 Eros. The periodic orbit family with binary period-doubling bifurcations is nearly circular, with almost zero inclination, and is reversed relative to the body of the asteroid 433 Eros. This implies that there are two stable regions separated by one unstable region for the motion around this asteroid. In addition, we found triple bifurcations which consist of two real saddle bifurcations and one period-doubling bifurcation. A periodic orbit family generated from an equilibrium point of asteroid 433 Eros has five bifurcations, which are one real saddle bifurcation, two tangent bifurcations, and two period-doubling bifurcations.
**Key words**: asteroidal system; periodic orbit families; bifurcations

## 1. Introduction

During the past few decades studies of the nonlinear dynamical behavior of a particle in the potential field of rotating highly irregular-shaped celestial bodies (hereafter called irregular bodies) have become more and more important and attractive to researchers ever since the increase in space exploration missions (Gong and Li 2015a, b; Jiang et al. 2015b; Jiang et al. 2016) to minor bodies and since the discoveries of large mass ratio binary or triple asteroids systems (Descamps et al. 2011; Durda et al. 2004). From the first flyby mission of 951 Gaspra by the Galileo spacecraft in 1991 (Belton et al. 1992), the first surrounding and soft-landing mission on 433 Eros by NASA's NEAR Shoemaker spacecraft (Veverka et al. 2001), to the first sampled return mission of 25143 Itokawa by the Hayabusa spacecraft of JAXA (Tsuchiyama et al. 2011), a large number of deep space explorations aimed at small bodies in solar system have been sent, e.g. ROSSETA, New Horizons, Dawn, Don Quijote, Hayabusa-II, and OSIRIS-REX. Theoretical research and actual missions can inform one another, which improves the understanding of the dynamical behavior of a particle affected by irregularly shaped bodies. This paper mainly discusses the multiple bifurcations phenomenon during the continuation of periodic orbits in the potential field of irregular bodies, such as asteroids, satellites of planets in the solar system, and comets. The results can be applied to studies of the dynamical behavior in the vicinity of irregularly shaped bodies, such as asteroids 216 Kleopatra, 433 Eros, and 6489 Golevka.

Periodic orbits around bodies with non-spherical potential fields have long been tracked. A lot of models have been developed to approximate the potential field in the vicinity of irregular bodies. Riaguas et al. (2001) and Gutierrez-Romero et al. (2004) investigated the stability of periodic orbit families around a massive straight segment as an approximation to elongated bodies. Hu and Scheeres (2004) expanded the potential field of an arbitrary asteroid up to the second degree and order, in order to check the existence and stability of equilibrium points and periodic motion. They found five basic families of periodic orbits in the gravity field up to the second degree and second order, and even further they discussed the dependence of orbital stability on the rotational rate (Hu and Scheeres 2008). Some over-simplified models, such as a homogeneous ellipsoid (Zhuravlev 1972), a rotating uniform mass density cube (Liu et al. 2011), and a rotating mass dipole connected by a massless bar (Zeng et al. 2015) have been used to discuss the existence, number, position, and stability of equilibrium points. Additionally, local periodic orbits near equilibrium points can be analysed according to properties of the local manifold. However, these over-simplified models may lose some of the essential characteristics of rotating non-central asymmetrical potential fields. For example, Scheeres et al. in 1996 failed to approximate the short-term energy effects of resonance in the vicinity of asteroids using the average method of perturbed Keplerian motion. And it is also trivial to realize that some of over-simplified models are far from the actual situation. The precise polyhedral method developed by Werner (1994) is a high-fidelity approximation of irregular bodies. Scheeres et al. (1996) computed families of periodic orbits around asteroid (4769) Castalia using Newton–Raphson iterations based on a homogeneous polyhedral model. Although this method produces a non-integrable system and increases computing cost compared to analytic models, the asymmetric form of the equations makes it nearly impossible to reduce the number of dimensions. Yet its detailed descriptions of the potential fields near the surface make it the first choice of a lot of scientists when investigating dynamical behavior. (Fahnestock and Scheeres 2008; Takahashi et al. 2013; Chanut et al. 2015)

The hierarchical grid searching method was proposed by Yu and Baoyin (2012) to search for periodic orbits. By introducing a definition of the Poincare section a new set of parameters is selected to describe a periodic orbit instead of the usual position and velocity parameters. Then one builds the mapping between the two sets of parameters. The new parameter space is a 2 layer grid with each layer having a 2-dimensional mesh. Trajectories are calculated on each grid point, and initial guesses of periodic orbits are given on the basis of the results; the precise value of the initial state of a periodic orbit is determined through further iteration. According to the continuity of periodic orbits every periodic orbit can be extended by varying the former initial state along the direction of the gradient of the generalized energy integral, i.e. the Jacobi constant, then making another iteration. This continuation leads to an intrinsic set of periodic orbits, the natural family, and its continuability is guaranteed by the fact that the periodic orbit can be represented locally by a power series of integral value. Meyer et al. (2009) gave the basic theory of this continuation. This method makes possible the systemic global search for 3D periodic orbits in the vicinity of irregular bodies.

Yu and Baoyin (2012) first used the hierarchical grid searching method and obtained 29 families of 3D periodic orbits near 216 Kleopatra. Ni et al. (2014) used their method and calculated 12 families of 3D periodic orbits with a solar perturbation in the vicinity of 433 Eros. In addition, Jiang et al. (2015a, b) calculated some periodic orbits near the surface of 6489 Golevka, 101955 Bennu, and Comet 1P / Halley. We need to establish a series of methods to classify periodic orbits and their families, so as to promote understanding of periodic orbits and their nonlinear dynamical behavior. Describing periodic orbits according to their shapes are the easiest and most intuitive method. However, it is not appropriate to reveal the mathematical mechanism, nor does it help when judging a particle's dynamic characteristics. Characteristic multipliers of a periodic orbit, that is, eigenvalues of the monodromy matrix of the periodic orbit, play an important role in classifying and understanding.

Orbits in the potential of the irregular body are motions in the 6-dimensional symplectic manifold. Riaguas et al. (1999) used characteristic multipliers and found several different periodic orbit families about a rotating segment. Using the symplectic eigenvalue theorem, Marsden and Ratiu (1999) presented 7 cases of orbits in a 4-dimensional symplectic manifold. Broucke and Elipe (2005) presented several periodic orbit families around a solid circular ring and used characteristic multipliers to analyze them. Yu and Baoyin (2012) fixed the two characteristic multipliers which equal 1 and used the classification results from Marsden and Ratiu (1999) to analyze the periodic orbits around an asteroid. Jiang et al. (2015a) applied the symplectic eigenvalue theorem to the 6-dimensional symplectic manifold and found that there are 34 cases for the orbits, out of which 13 cases are periodic. In addition, they presented a theory on the global motion around irregular bodies, including the structure of submanifolds, bifurcations of periodic orbit families, and different topological transfers for periodic orbits of bifurcations. After that, Yu et al. (2015) used the 13 classifications of periodic orbits about irregular bodies from Jiang et al. (2015a) and calculated several periodic orbit families around asteroid 243 Ida. Jiang et al. (2015b) furthermore discussed several different kinds of resonant periodic orbits around asteroid 216 Kleopatra, which have different resonant ratios and topological cases.

Research on bifurcations of periodic orbits in the continuation family has progressed from analysing a single orbit to dynamic varying processes (Jiang et al. 2015a). All four theoretical types of bifurcations exist in the actual situation. However, further research has recognized multiple bifurcations during continuation of the initial periodic orbits of 433 Eros. Two or three different bifurcations or two of the same bifurcations appear when the Jacobian constants of periodic orbits continuously vary. There are been several different kinds of multiple bifurcations in the orbit families near 433 Eros. The mechanism and conditions of bifurcation that appear in combination during continuation are clearly revealed by the analysis.

This paper is organized as follows. The characteristic multipliers and topological classifications of periodic orbits around an irregular body are discussed in Section 2. The properties of the four kinds of basic bifurcations in periodic orbit families are presented in detail. Multiple bifurcations in periodic orbit families in the potential of an irregular body are investigated in Section 3. All the multiple bifurcations in periodic orbit families consist of four kinds of basic bifurcations.

We found binary period-doubling bifurcations and binary tangent bifurcations in periodic orbit families around asteroid 433 Eros. The periodic orbit family which shows the binary period-doubling bifurcations is nearly circular, has zero inclinations, and is reversed relative to the body of the asteroid; the result shows that there are two stable regions divided by one unstable region around asteroid 433 Eros. Triple bifurcations in periodic orbit families consisting of two real saddle bifurcations and one period-doubling bifurcation have been found; after the first real saddle bifurcation, the characteristic multipliers move from the positive half plane to the negative half plane. Multiple bifurcations in periodic orbit families which consist of one real saddle bifurcation, two tangent bifurcations, and two period-doubling bifurcations are found; this periodic orbit family is continued from an equilibrium point of asteroid 433 Eros.

## 2. Characteristic Multipliers and Basic Bifurcations of Periodic Orbits

### 2.1 Characteristic Multipliers

Denoting $r$ as the body-fixed vector from the celestial body's centre of mass to the particle, $\omega$ as the rotational angular velocity of the celestial body relative to the inertial system, and $U(r)$ as the gravitational potential of the body at $r$, then the effective potential (Scheeres et al. 1996, 1998; Scheeres 2012; Jiang and Baoyin 2014; Jiang 2015; Yu et al. 2015) can be defined by Equation (1):

$$V(r) = -\frac{1}{2}(\omega \times r) \cdot (\omega \times r) + U(r). \tag{1}$$

For irregular-shaped celestial bodies $U(r)$ is calculated by polyhedral models using radar observation data (Werner and Scheeres 1997). Specifically, Equations (2) and (3) reflect this:

$$U = \frac{1}{2}G\sigma \sum_{e \in edges} r_e \cdot E_e \cdot r_e \cdot L_e - \frac{1}{2}G\sigma \sum_{f \in faces} r_f \cdot F_f \cdot r_f \cdot \omega_f, \tag{2}$$

and

$$\nabla U = -G\sigma \sum_{e \in edges} E_e \cdot r_e \cdot L_e + G\sigma \sum_{f \in faces} F_f \cdot r_f \cdot \omega_f, \tag{3}$$

where $G$ denotes the gravitational constant, $\sigma$ represents the celestial body's bulk density, $r_e$ and $r_f$ are body-fixed vectors from the particle at $r$ to edge $e$ and face $f$, respectively; while $E_e$ and $F_f$ are geometric parameters of the edges and faces, respectively. $L_e$ means the integration factor of the particle position and edge $e$. $\omega_f$ is the solid angle of the face $f$ relative to the particle.

Then the equation of motion (Scheeres 2012; Zotos 2015a, b) may be written as Equation (4):

$$\ddot{r} + 2\boldsymbol{\omega} \times \dot{r} + \frac{\partial V(r)}{\partial r} = \mathbf{0}, \qquad (4)$$

with $V$ as the effective potential.

We first define the conjugate variables $p = \boldsymbol{\omega} \times r + \dot{r}$ and $q = r$, then introduce the state vector $x = [p \quad q]^{\mathrm{T}}$. Since the Hamiltonian (Zotos 2015a, b) is given by Equation (5):

$$H = -\frac{1}{2} p \cdot p + p \cdot \dot{q} + U(q), \qquad (5)$$

then the dynamical equation can be written as a symplectic form given by Equation (6):

$$\dot{x} = f(x) = J \nabla H(x), \qquad (6)$$

where $J$ has the form given by Equation (7):

$$J = \begin{bmatrix} O & -I \\ I & O \end{bmatrix}, \qquad (7)$$

and $O$ and $I$ are $3 \times 3$ zero and identity matrices, respectively. With the symplectic quality of $J$, that is, $J^{-1} = -J$, the dynamical equation can be rewritten as Equation (8):

$$J \dot{x} + \nabla H(x) = \mathbf{0}. \qquad (8)$$

If $S_p(T)$ denotes the set of periodic orbits with period $T$, then $\forall\, p \in S_p(T)$, considering the $6 \times 6$ matrix $\partial f / \partial x$, the state transition matrix for the periodic orbit can be written as Equation (9):

$$\boldsymbol{\Phi}(t) = \int_0^t \frac{\partial \boldsymbol{f}}{\partial \boldsymbol{x}}(p(\tau)) \mathrm{d}\tau. \tag{9}$$

The monodromy matrix of the periodic orbit $p \in S_p(T)$ is given by Equation (10):

$$\boldsymbol{M} = \boldsymbol{\Phi}(T). \tag{10}$$

This matrix is symplectic. Some important properties of $\boldsymbol{M}$ are: (i) if $\lambda$ is an eigenvalue of the matrix, then so are $\lambda^{-1}$, $\bar{\lambda}$ and $\bar{\lambda}^{-1}$; namely, all eigenvalues have one of the following forms: 1, −1, $e^{\pm\alpha}(\alpha \in (0, 1))$, $-e^{\pm\alpha}(\alpha \in (-1, 0))$, $e^{\pm i\beta}(\beta \in (0, \pi))$, and $e^{\pm\sigma \pm i\tau}(\sigma > 0, \tau \in (0, \pi))$. (ii) det$(\boldsymbol{M}) = 1$, so 0 is not an eigenvalue of $\boldsymbol{M}$, thus, there is at least one characteristic multiplier equal to 1, and the algebraic multiplicities of multipliers ±1 must be even. Eigenvalues of this matrix are characteristic multipliers of the periodic orbit, also called Floquet multipliers, which are fixed to a specified periodic orbit $p$.

**2.2 Topological structures related with bifurcations**

Jiang et al. (2015a) have derived and summarized all of the classifications of an orbit in the potential field of a rotating celestial body. In our research of orbit family continuation, 10 out of the 13 topological structures for periodic orbits appeared and 9 of those can be found before and after bifurcations. The characteristics of these cases are presented in Table 1. To facilitate comparison the names of the cases given in Jiang et al. (2015a) are adopted. Cases P1-P7 represent the purely periodic cases. Case PDRS1 represents the periodic and degenerate real saddle case. Cases PPD3 and PPD4 represent the periodic and period-doubling cases.

Cases P1-P4 are 4 basic topological structures that occur during periodic orbit family continuation. These 4 basic topological structures may change according to specific principles via other cases, which is called bifurcation. In this paper Cases 5, 6, PDRS1, PPD3 and PPD4 are bridges between the basic topological structures. Paths of bifurcation during periodic orbits continuation will be introduced in the following section.

Table 1. The topological cases related with bifurcations (Jiang et al. 2015a)

| Cases | Structure of characteristic multipliers |
| --- | --- |
| P1 | $\gamma_j$ ($\gamma_j = 1$; $j = 1, 2$) and $e^{\pm\sigma \pm i\tau}$ ($\sigma > 0$, $\tau \in (0, \pi)$) |
| P2 | $\gamma_j$ ($\gamma_j = 1$; $j = 1, 2$) and $e^{\pm i\beta_j}$ ($\beta_j \in (0, \pi)$; $j = 1, 2 \mid \beta_1 \neq \beta_2$) |
| P3 | $\gamma_j$ ($\gamma_j = 1$; $j = 1, 2$) and $e^{\pm\alpha_j}$ ($\alpha_j \in (0, 1)$; $j = 1, 2 \mid \alpha_1 \neq \alpha_2$) or $-e^{\pm\alpha_j}$ ($\alpha_j \in (-1, 0)$; $j = 1, 2 \mid \alpha_1 \neq \alpha_2$) |
| P4 | $\gamma_j$ ($\gamma_j = 1$; $j = 1, 2$), $e^{\pm i\beta}$ ($\beta \in (0, \pi)$) and $e^{\pm\alpha}$ ($\alpha \in (0, 1)$) or $-e^{\pm\alpha}$ ($\alpha \in (-1, 0)$) |
| P5 | $\gamma_j$ ($\gamma_j = 1$; $j = 1, 2, 3, 4$) and $e^{\pm i\beta}$ ($\beta \in (0, \pi)$) |
| P6 | $\gamma_j$ ($\gamma_j = 1$; $j = 1, 2, 3, 4$) and $e^{\pm\alpha}$ ($\alpha \in (0, 1)$) or $-e^{\pm\alpha}$ ($\alpha \in (-1, 0)$) |
| P7 | $\gamma_j$ ($\gamma_j = 1$; $j = 1, 2, 3, 4, 5, 6$) |
| PDRS1 | $\gamma_j$ ($\gamma_j = 1$; $j = 1, 2$) and $e^{\pm\alpha_j}$ ($\alpha_j \in (0, 1)$; $j = 1, 2 \mid \alpha_1 = \alpha_2$) or $-e^{\pm\alpha_j}$ ($\alpha_j \in (-1, 0)$; $j = 1, 2 \mid \alpha_1 = \alpha_2$) |
| PPD3 | $\gamma_j$ ($\gamma_j = 1$; $j = 1, 2$), $\gamma_j$ ($\gamma_j = -1$; $j = 3, 4$) and $e^{\pm i\beta}$ ($\beta \in (0, \pi)$) |
| PPD4 | $\gamma_j$ ($\gamma_j = 1$; $j = 1, 2$), $\gamma_j$ ($\gamma_j = -1$; $j = 3, 4$) and $e^{\pm\alpha}$ ($\alpha \in (0, 1)$) or $-e^{\pm\alpha}$ ($\alpha \in (-1, 0)$) |

## 2.3 Four kinds of Basic Bifurcations

There are altogether 4 kinds of bifurcations of periodic orbits in the potential field of irregular shaped celestial bodies, including period-doubling bifurcations, tangent bifurcations, real saddle bifurcations, and Neimark-Sacker bifurcations (Jiang et al. 2015a). Yu et al. (2015) also applied the theory from Jiang et al. (2015a) to analyze the continuation of periodic orbits around asteroid 243 Ida; however, they only discussed the topological transfers of periodic orbits and didn't further analyze the bifurcations. Period-doubling bifurcations occur when the orbits have characteristic multipliers that cross $-1$, and the motion of the characteristic multipliers causes them to leave the unit circle for the real axis (or from real axis to unit circle, contrarily); the multiplicities of $-1$ are 2 or 4. Tangent bifurcations occur when the orbits have characteristic multipliers that cross 1, and the motion of the characteristic multipliers causes them to leave the unit circle for the real axis (or from real axis to unit circle, contrarily); the multiplicities of 1 are 4 or 6. Real saddle bifurcations occur when the orbits have characteristic multipliers that cross the real axis, and the motion of the characteristic multipliers causes them to leave the real axis (or be limited to the real axis, contrarily). Neimark-Sacker bifurcations occur when the orbits have two equal characteristic multipliers $e^{i\beta}$ or $e^{-i\beta}$ ($\beta \in (0, \pi)$), and the motion of the characteristic multipliers causes them to leave the unit circle (or be limited on the unit circle, contrarily). In addition, pseudo bifurcations happen if the multipliers cross with no change in motion. Our research of multiple bifurcations of periodic orbit continuation brings out three of these four basic bifurcations with the following paths.

The period-doubling bifurcations (Jiang et al. 2015b) for periodic orbit families have the following two paths:

Period-doubling Bifurcation I. The transfer of the topological case follows Case P4 → Case PPD3 → Case P2 or Case P2 → Case PPD3 → Case P4.

Period-doubling Bifurcation II. The transfer of the topological case follows Case P4 → Case PPD4 → Case P3 or Case P3 → Case PPD4 → Case P4.

The pseudo period-doubling bifurcations (Jiang et al. 2015b) for periodic orbit families have the following four paths:

Pseudo Period-doubling Bifurcation I. The transfer of the topological case follows Case P4 → Case PPD3 → Case P4.

Pseudo Period-doubling Bifurcation II. The transfer of the topological case follows Case P2 → Case PPD3 → Case P2.

Pseudo Period-doubling Bifurcation III. The transfer of the topological case follows Case P4 → Case PPD4 → Case P4.

Pseudo Period-doubling Bifurcation IV. The transfer of the topological case follows Case P3 → Case PPD4 → Case P3.

The tangent bifurcations for periodic orbit families have the following two paths:

Tangent Bifurcation I. The transfer of the topological case follows Case P2 → Case P5 → Case P4 or Case P4 → Case P5 → Case P2.

Tangent Bifurcation II. The transfer of the topological case follows Case P4 → Case P6 → Case P3 or Case P3 → Case P6 → Case P4.

The pseudo tangent bifurcations for periodic orbit families have the following

four paths:

Pseudo Tangent Bifurcations I. The transfer of the topological case follows Case P2 → Case P5 → Case P2.

Pseudo Tangent Bifurcations II. The transfer of the topological case follows Case P3 → Case P6 → Case P3.

Pseudo Tangent Bifurcations III. The transfer of the topological case follows Case P4 → Case P5 → Case P4.

Pseudo Tangent Bifurcations IV. The transfer of the topological case follows Case P4 → Case P6 → Case P4.

The real saddle bifurcations for periodic orbit families have the following path:

Real Saddle Bifurcation I. The transfer of the topological case follows Case P1 → Case PDRS1 → Case P3 or Case P3 → Case PDRS1 → Case P1.

The pseudo real saddle bifurcations for periodic orbit families have the following two paths:

Pseudo Real Saddle Bifurcation I. The transfer of the topological case follows Case P1 → Case PDRS1 → Case P1.

Pseudo Real Saddle Bifurcation II. The transfer of the topological case follows Case P3 → Case PDRS1 → Case P3.

It is important to distinguish pseudo bifurcations from real saddle bifurcations. Thus it is necessary to check topological cases over a relatively large range during continuations.

## 3. Multiple Bifurcations of Periodic Orbit Families

The periodic orbit families in the potential of an irregular body not only have four kinds of bifurcations and pseudo bifurcations, but also have multiple bifurcations. First we investigate the possible multiple bifurcations; second we discuss four types of multiple bifurcations of periodic orbit families which are found about asteroid 433 Eros.

## 3.1 Multiple Bifurcations Consisting of Four Kinds of Basic Bifurcations

The periodic orbit family is generated after the continuation of periodic orbits. In a single periodic orbit family, there may not have been only one bifurcation. That is to say, multiple bifurcations consisting of the four kinds of basic bifurcations may be found in the periodic orbit family.

There exist 10 kinds of binary bifurcations. For instance, two period-doubling bifurcations occur in a single periodic orbit family during the continuation; these are the binary period-doubling bifurcations. Similarly, binary tangent bifurcations also exist, as well as the binary Neimark-Sacker bifurcation, and the binary real saddle bifurcation. For example, the paths of topological cases of periodic orbits Case P2 → Case PPD3 → Case P4 → Case PPD3 →Case P2 lead to the binary period-doubling bifurcations. The paths of these two period-doubling bifurcations have opposite directions.

In addition, if the two bifurcations in the binary bifurcations are different, then there are six different cases of the binary bifurcations, i.e. the period-doubling bifurcation and the tangent bifurcation, the period-doubling bifurcation and the Neimark-Sacker bifurcation, the period-doubling bifurcation and the real saddle bifurcation, the tangent bifurcation and the Neimark-Sacker bifurcation, the tangent bifurcation and the real saddle bifurcation, and the Neimark-Sacker bifurcation and the real saddle bifurcation.

**3.2 Binary Period-Doubling Bifurcations**

Binary period-doubling bifurcations exist in the same periodic orbit family. Figure 1 shows a periodic orbit family about asteroid 433 Eros which has binary period-doubling bifurcations. In the three panels in Figure 1 the left Figure shows the periodic orbits in the xyz frame while the right Figure shows the periodic orbits in the xy plane. The Jacobian constants of the periodic orbits vary from 25.79 $m^2s^{-2}$ to 20.79 $m^2s^{-2}$, and the periods of the periodic orbits vary from $1.1703\times10^4$ s to $0.9729\times10^4$ s. The first period-doubling bifurcation occurs when the Jacobian constant equals 23.19 $m^2s^{-2}$ and the period equals $1.0777\times10^4$ s. The second period-doubling bifurcation occurs when the Jacobian constant equals 22.79 $m^2s^{-2}$ and the period equals $1.0620\times10^4$ s. The Figure (1a) shows the periodic orbits before the first period-doubling bifurcation while the Figure (1c) shows the periodic orbits after the second period-doubling bifurcation. Figure (1b) shows the periodic orbits between the first and the second period-doubling bifurcations. Figure 2 illustrates this continuation process using the relationship between the Jacobian constant and the periodicity of the orbits. Figure 3 shows the topological transfer paths in the periodic orbit family with the binary period-doubling bifurcations.

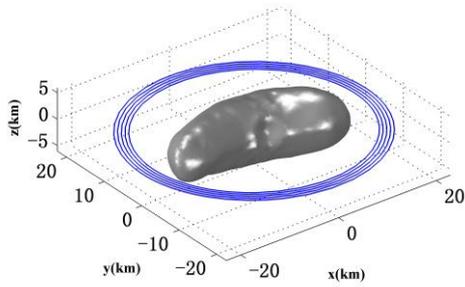 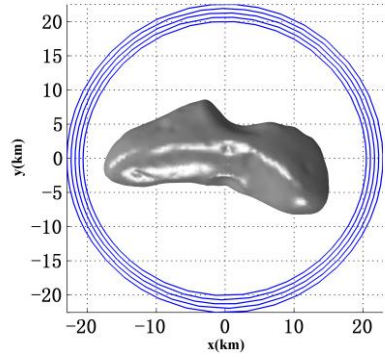

a

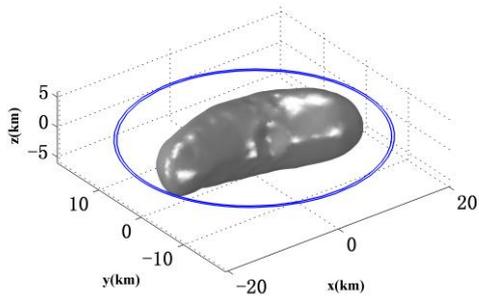 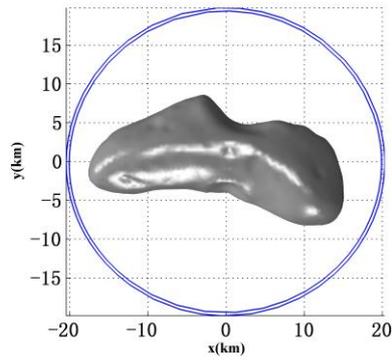

b

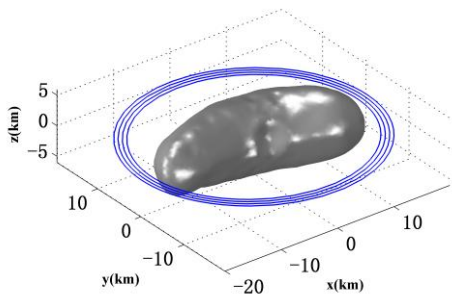 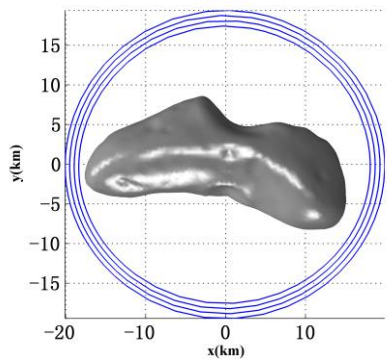

c

Figure 1. A periodic orbit family about asteroid 433 Eros and its binary period-doubling bifurcations: (a) shows the continuation of periodic orbits before the first period-doubling bifurcation, (b) shows the continuation of periodic orbits after the first period-doubling bifurcation and before the second period-doubling bifurcation, (c) shows the continuation of periodic orbits after the second period-doubling bifurcation

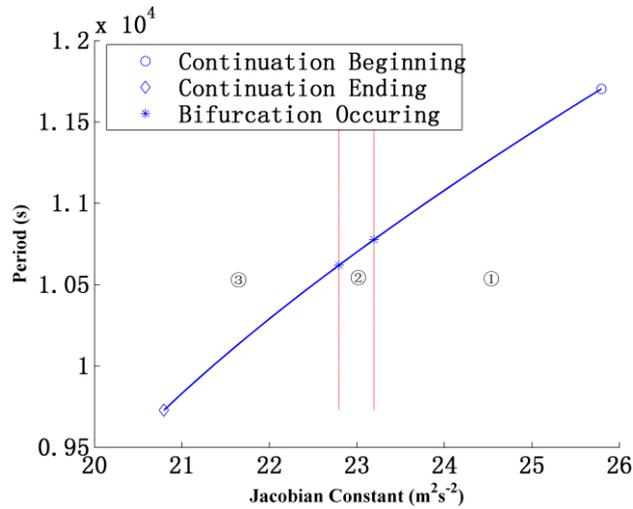

Figure 2. Illustration of the periodic orbit family continuation containing binary period-doubling bifurcations, including the period's variation of orbits with the changing Jacobian constant and their relationship when bifurcations occur. Topological cases in zone 1, 2, and 3 are Case P2, Case P4 and Case P2, respectively.

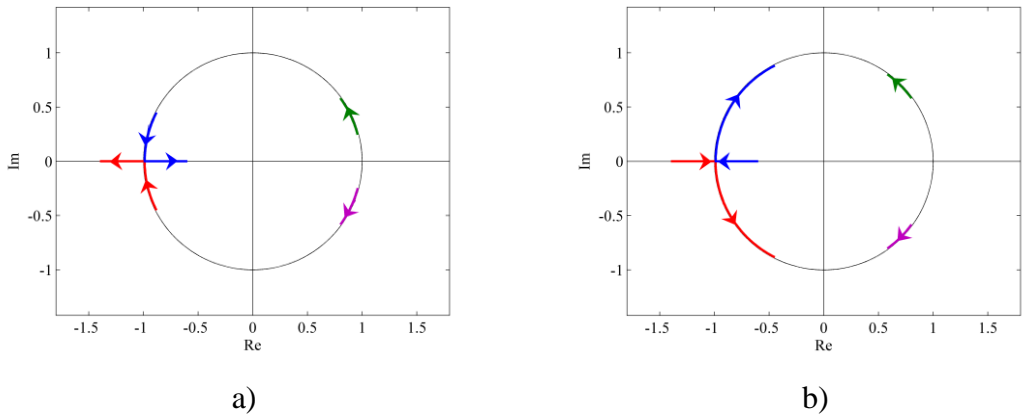

a)            b)

Figure 3. The topological transfer paths of the binary period-doubling bifurcations: a) shows the path of the first period-doubling bifurcation while b) shows the path of the second period-doubling bifurcation.

Before the first period-doubling bifurcation, the topological cases of the periodic orbits in the periodic orbit family belong to Case P2. During the continuation the topological case of the periodic orbit belongs to Case PPD3 when the first bifurcation occurs. After the bifurcation the topological cases of the periodic orbits belong to Case P4. The stability of the periodic orbits changes from stable to unstable.

Before the second period-doubling bifurcation the periodic orbits' topological cases belong to Case P4, and these periodic orbits are all unstable. The continuation goes on and the second period-doubling bifurcation occurs when the Jacobian constant equals 23.19 $m^2s^{-2}$, as we said before. When the second period-doubling bifurcation appears, the topological case of the periodic orbit is Case PPD3. The sojourn time of the topological case at Case PPD3 is zero. In other words, there is only one periodic orbit which belongs to Case PPD3 during the second bifurcation. After the second bifurcation the topological case of the periodic orbit is Case P2, and the periodic orbits become stable.

This periodic orbit family is nearly circular, with zero inclination, and is reversed relative to the body of the asteroid 433 Eros. Jiang et al. (2015b) calculated a family of periodic orbits around asteroid 216 Kleopatra, and found the orbit family is stable, nearly circular, with zero inclination, and is reversed relative to the body of the asteroid 216 Kleopatra. Their orbit family has no bifurcations, only the pseudo period-doubling bifurcation; the characteristic multipliers in the unit circle collide at -1, and pass through each other after the collision. They conclude that there exists a stable region around the large size ratio triple asteroid 216 Kleopatra and give an explanation of the orbit stability of the two moonlets, Alexhelios and Cleoselene. Our orbit family around asteroid 433 Eros has two period-doubling bifurcations. In the first period-doubling bifurcation the characteristic multipliers in the unit circle collide at -1, and leave the unit circle after the collision. Then the orbits in the periodic orbit family become unstable. After the second period-doubling bifurcation orbits in the periodic orbit family become stable again. This result means that there is an unstable region between the two stable regions, which causes the stable region to narrow.

**3.3 Binary Tangent Bifurcation**

Binary tangent bifurcations in the same periodic orbit family continuation are also possible. Figure 4 shows a binary tangent bifurcation process about a periodic orbit family of asteroid 433 Eros. In the three panels in Figure 3 the periodic orbits in the xyz frame are shown on the left while the right Figures show the periodic orbits in the xy plane. The Jacobian constants of the periodic orbits vary from $-26.99$ $m^2s^{-2}$ to $-50.75$ $m^2s^{-2}$, and the periods of the periodic orbits vary from $3.7427\times10^4$ s to $2.5477\times10^4$ s. The first tangent bifurcation occurs when the Jacobian constant equals $-44.39$ $m^2s^{-2}$ and the period equals $3.6901\times10^4$ s. The second tangent bifurcation occurs when the Jacobian constant equals $-46.55$ $m^2s^{-2}$ and the period equals $3.1778\times10^4$ s. Figure (4a) shows the periodic orbits before the first tangent bifurcation while Figure (4c) shows the periodic orbits after the second tangent bifurcation. Figure (4b) shows the periodic orbits between these two tangent bifurcations. Figure 5 illustrates this continuation process using the relationship between the Jacobian constant and periodicity of the orbits. Figure 6 shows the topological transfer paths in this periodic orbit family with the binary tangent bifurcations.

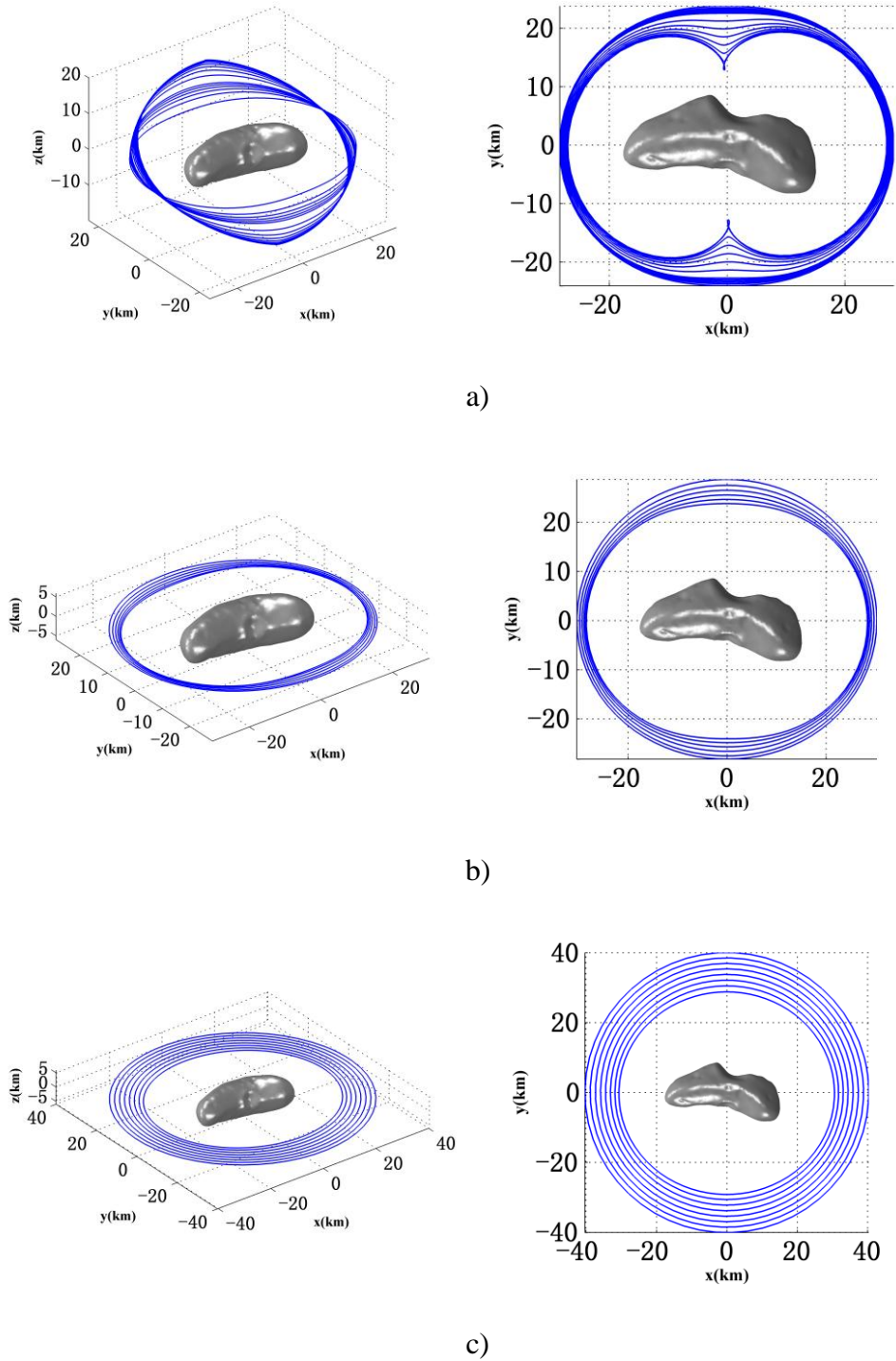

Figure 4. A periodic orbit family about asteroid 433 Eros and its binary tangent bifurcations: (a) shows the continuation of periodic orbits before the first tangent bifurcation; (b) shows the continuation of periodic orbits after the first tangent bifurcation and before the second tangent bifurcation; and (c) shows the continuation of periodic orbits after the second tangent bifurcation.

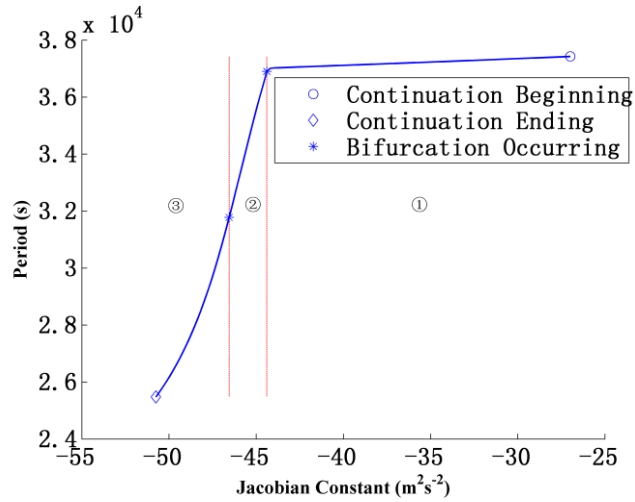

Figure 5. Illustration of the periodic orbit family continuation containing binary tangent bifurcations, including the period's variation of orbits with a changing Jacobian constant and their relationship when bifurcations occur. Topological cases in zone 1, 2, and 3 are Case P3, Case P4, and Case P2, respectively.

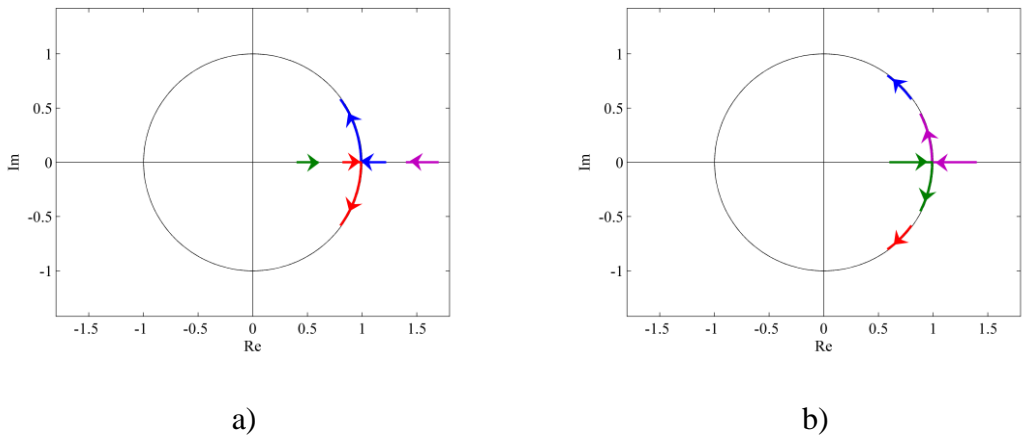

a)　　　　　　　　　　　　　　b)

Figure 6. The topological transfer paths of the binary tangent bifurcations: a) shows the path of the first tangent bifurcation, and b) shows the path of the second tangent bifurcation.

Before the first tangent bifurcation the topological cases of the periodic orbits in the periodic orbit family belong to Case P3. During the continuation the topological case of the periodic orbit belongs to Case P6 when the first bifurcation occurs. After the bifurcation the topological cases of the periodic orbits belong to Case P4. The periodic orbits remain unstable. From Figure 5 one can see that after the first tangent bifurcation, the rate of change for the period and Jacobian constant for periodic orbits obviously vary. The trajectory of the periods and the Jacobian constants of the periodic orbit family have a cusp when the first tangent bifurcation occurs. In other words, the trajectory is not smooth when the first tangent bifurcation occurs.

Before the second tangent bifurcation the periodic orbits' topological cases belong to Case P4, and these periodic orbits are all unstable. The continuation goes on and the second tangent bifurcation occurs when the Jacobian constant equals the forementioned value $-46.55$ $m^2s^{-2}$. When the second tangent bifurcation appears, the topological case of the periodic orbit is Case P5. The sojourn time of topological case at Case P5 is zero. In other words, there is only one periodic orbit which belongs to Case P5 during the second bifurcation. After the second bifurcation the topological case of the periodic orbit is Case P2, and the periodic orbits become stable.

## 3.4 Triple Bifurcations consisting of Two Real Saddle Bifurcation and a Period-Doubling Bifurcation

The combination of real saddle bifurcation and period-doubling bifurcation in the same periodic orbit family also exist. A periodic orbit family about asteroid 433 Eros which combines a binary real saddle bifurcation and period-doubling bifurcations is presented in Figure 7. Similar to the Figures in Section 3.2, all the left Figures in Figure 7 show the periodic orbits in the xyz frame while the right ones show the periodic orbits in the xy plane. The Jacobian constants of the periodic orbits vary from $-35.18$ $m^2s^{-2}$ to $-22.78$ $m^2s^{-2}$, and the periods of the periodic orbits vary from $3.7079 \times 10^4$ s to $3.5651 \times 10^4$ s. The first real saddle bifurcation occurs when the Jacobian constant equals $-33.98$ m$^2$s$^{-2}$ and the period equals $3.7079\times 10^4$ s. The second real saddle bifurcation occurs when the Jacobian constant equals $-29.78$ m$^2$s$^{-2}$ and the period equals $3.6992\times 10^4$ s. The period-doubling bifurcation occurs when the Jacobian constant equals $-27.58$ m$^2$s$^{-2}$ and the period equals $3.6825\times 10^4$ s. Figure (7a) shows the periodic orbits before the first real saddle bifurcation; figure (7b) shows the periodic orbits after the first real saddle bifurcation and before the second one; figure (7c) shows the periodic orbits after the second real saddle bifurcation and before the period-doubling bifurcation. figure (7d) shows the periodic orbits after the period-doubling bifurcations. Figure 8 illustrates this continuation process using the relationship between the Jacobian constant and periodicity of the orbits. Figure 9 shows the topological transfer paths in this periodic orbit family with the triple bifurcations.

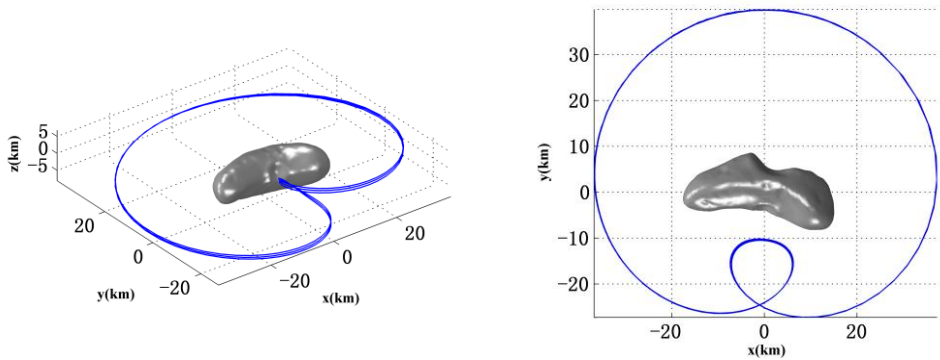

a)

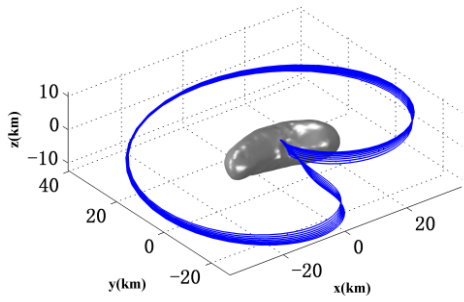
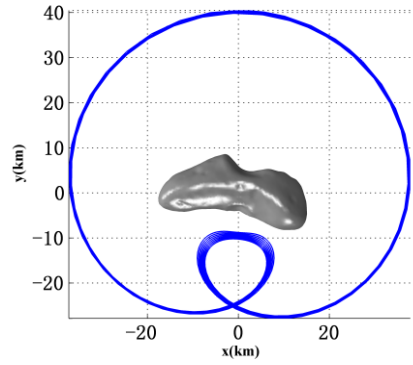

b)

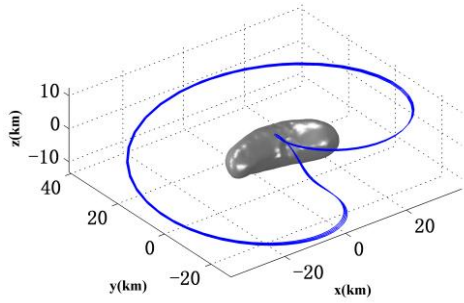
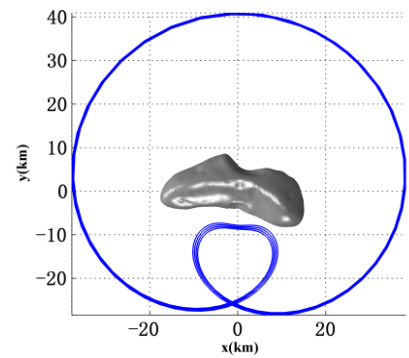

c)

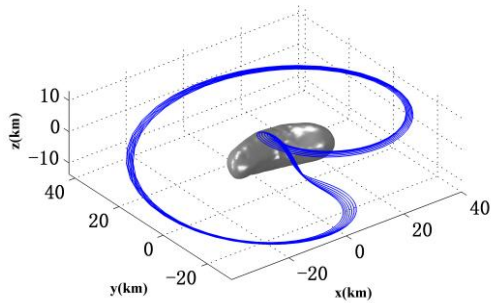
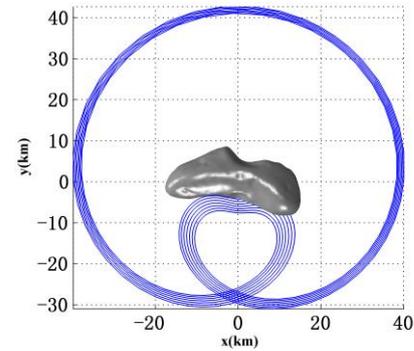

d)

Figure 7. A periodic orbit family about asteroid 433 Eros and its triple bifurcations consisted of binary real saddle and period-doubling bifurcation: (a) shows the continuation of periodic orbits before the first real saddle bifurcation; (b) shows the continuation of periodic orbits after the first real saddle bifurcation and before the second real saddle bifurcation; (c) shows the continuation of periodic orbits after the second real saddle bifurcation and before the period-doubling bifurcation; and (d) shows the continuation of periodic orbits after the period-doubling bifurcation.

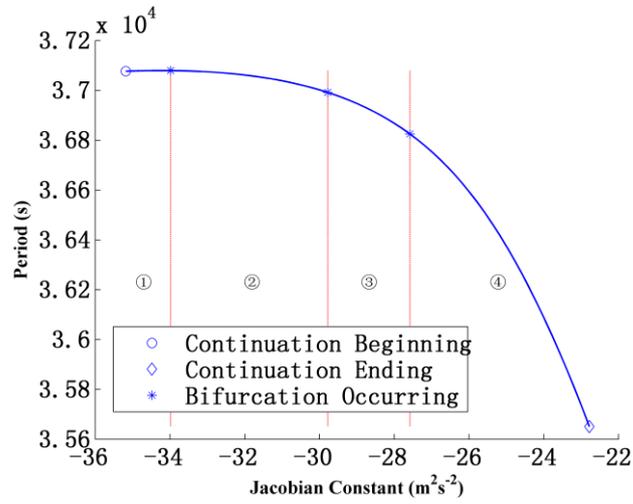

Figure 8. Illustration of the periodic orbit family continuation containing triple bifurcations, including the period's variation of orbits with a changing Jacobian constant and their relationship when bifurcations occurring. Topological cases in zone 1, 2, 3, and 4 are Case P3, Case P1, Case P3 and Case P2, respectively.

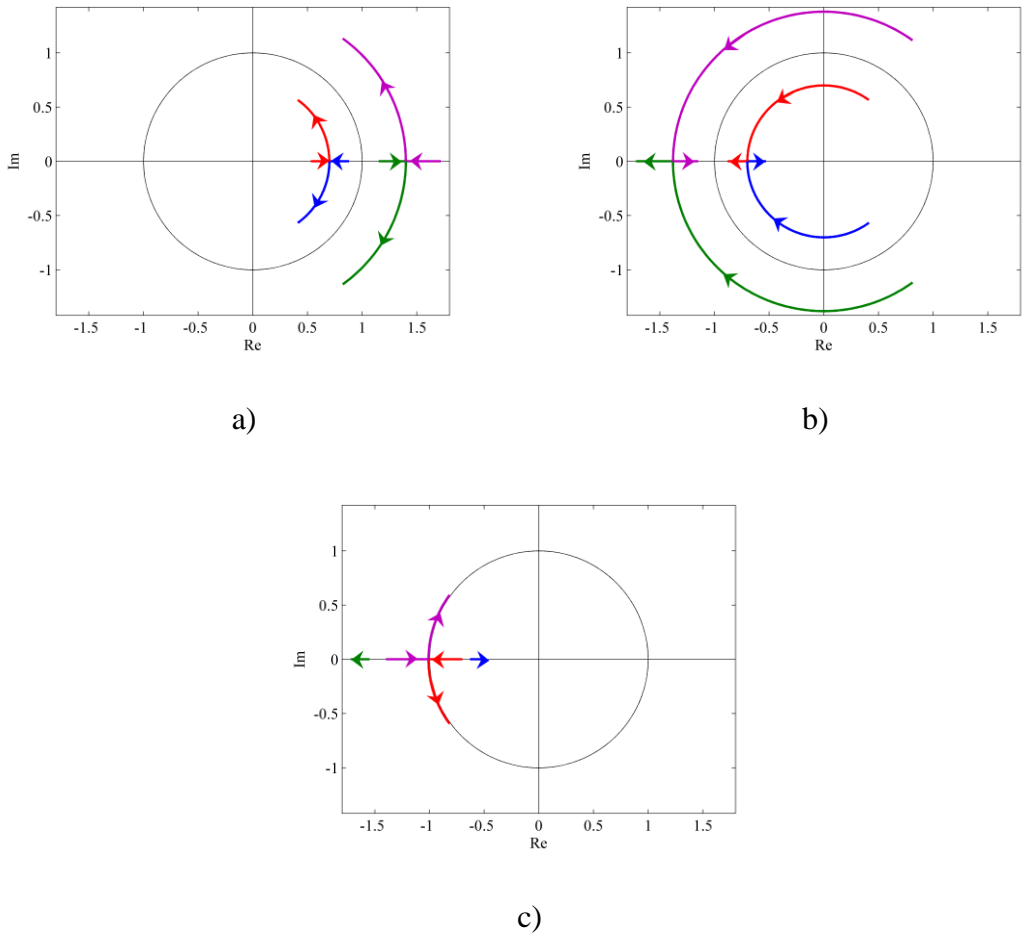

Figure 9. The topological transfer paths of the triple bifurcations: a) shows the path of the first real saddle bifurcation, b) shows the path of the second real saddle bifurcation, and c) shows the path

of the period-doubling bifurcation.

Before the first real saddle bifurcation the topological cases of the periodic orbits in the periodic orbit family belong to Case P3. During the continuation the topological case of the periodic orbit belongs to Case PDRS1 when the Jacobian constant equals −33.98 $m^2s^{-2}$ and the first bifurcation occurs. The sojourn time of the topological case at Case PDRS1 is zero. As mentioned before there is only one periodic orbit belonging to Case PDRS1 during the bifurcation. After the bifurcation the topological cases of the periodic orbits belong to Case P1. The periodic orbits remain unstable. From Figure 9 one can see that after the first bifurcation and before the second bifurcation, the characteristic multipliers move from the positive half plane to the negative half plane. Then after the second real saddle bifurcation the period-doubling bifurcation occurs. If the characteristic multipliers don't move from the positive half plane to the negative half plane, the period-doubling bifurcation will never occur during the continuation.

Before the second real saddle bifurcation, the periodic orbits' topological cases belong to Case P1, and these periodic orbits are all unstable. The continuation goes on and the second real saddle bifurcation occurs when the Jacobian constant equals −29.78 $m^2s^{-2}$, as we said before. When the second period-doubling bifurcation appears, the topological case of the periodic orbit is Case PDRS1. The topological cases of the periodic orbits after the second bifurcation revert back to Case P3, and these orbits are still unstable.

After the second real saddle bifurcation and before the period-doubling bifurcation, the topological cases of the periodic orbits in the periodic orbit family belong to Case P3. During the continuation the topological case of the periodic orbit belongs to Case PPD4 when the Jacobian constant equals $-27.58$ m$^2$s$^{-2}$ and the third bifurcation occurs. As mentioned before there is only one periodic orbit belonging to Case PPD4 during the bifurcation. After the bifurcation the topological cases of the periodic orbits belong to Case P2. The periodic orbits become stable.

**3.5 Multiple Mixed Bifurcations**

The continuations of periodic orbits around the equilibrium points (Chanut et al. 2015; Elipe and Lara 2003; Elipe and Riaguas 2003; Liu et al. 2011; Jiang et al. 2014, 2016) have no bifurcation if the amplitudes (Jiang 2015) of the periodic orbits are sufficiently small. However, if the amplitudes of the periodic orbits increase, bifurcations of periodic orbits do exist, even multiple mixed bifurcations may occur. Figure 10 is an example of the extraordinary multiple mixed bifurcations contained in the continuation of an ordinary periodic orbit family in the vicinity of an equilibrium point of asteroid 433 Eros. Similar to previous sections all left Figures in Figure 10 show the periodic orbits in the xyz frame while the right ones show the periodic orbits in the xz plane. Figure 12 shows the topological transfer paths in this periodic orbit family with the multiple mixed bifurcations.

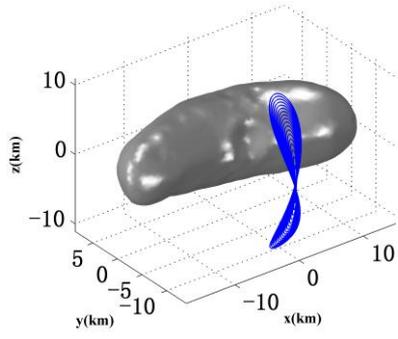
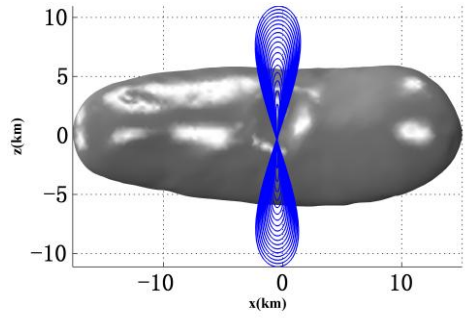

a)

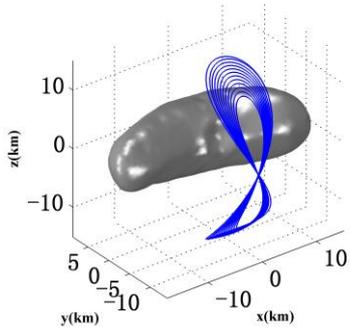
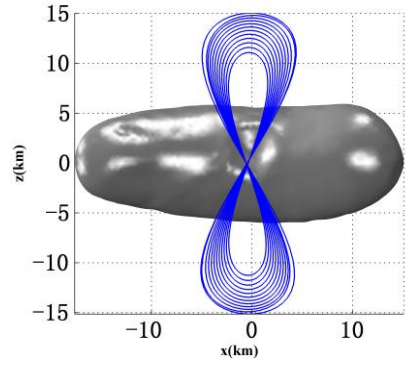

b)

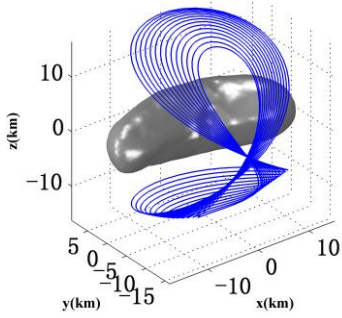
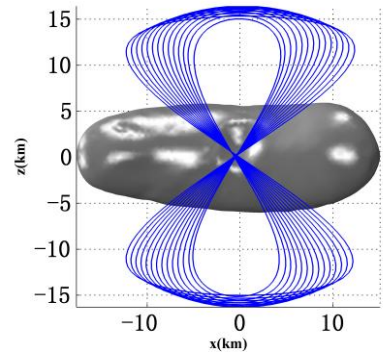

c)

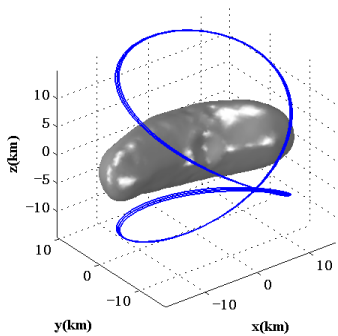
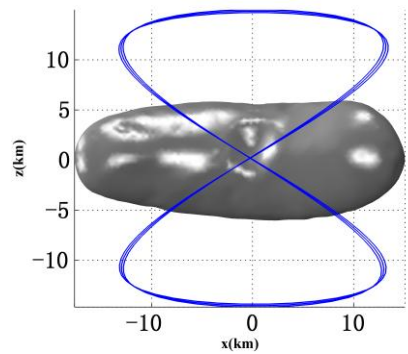

d)

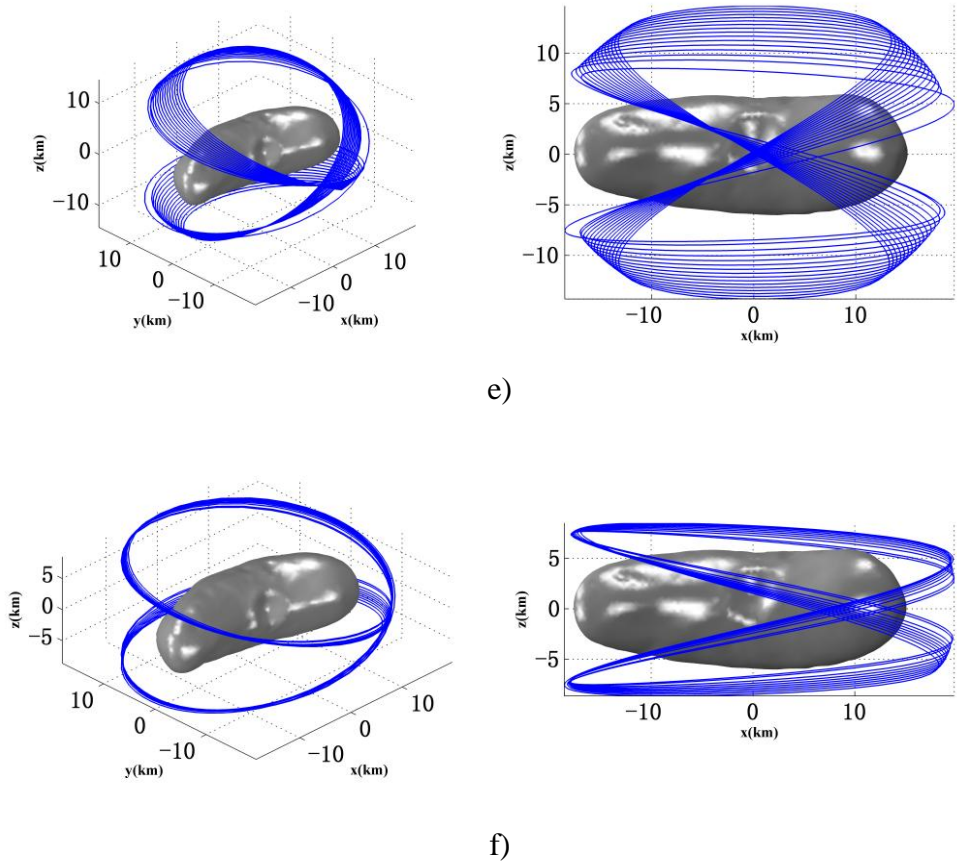

e)

f)

Figure 10. A periodic orbit family about asteroid 433 Eros and its multiple mixed bifurcations: (a) shows the continuation of periodic orbits before the real saddle bifurcation; (b) shows the continuation of periodic orbits after the real saddle bifurcation and before the first tangent bifurcation; (c) shows the continuation of periodic orbits between the first tangent bifurcation and the first periodic-doubling bifurcation; (d) shows the continuation of periodic orbits between two periodic-doubling bifurcations; (e) shows the continuation of periodic orbits after the second periodic-doubling bifurcation and before the second tangent bifurcation; and (f) shows the continuation of periodic orbits after the second tangent bifurcation.

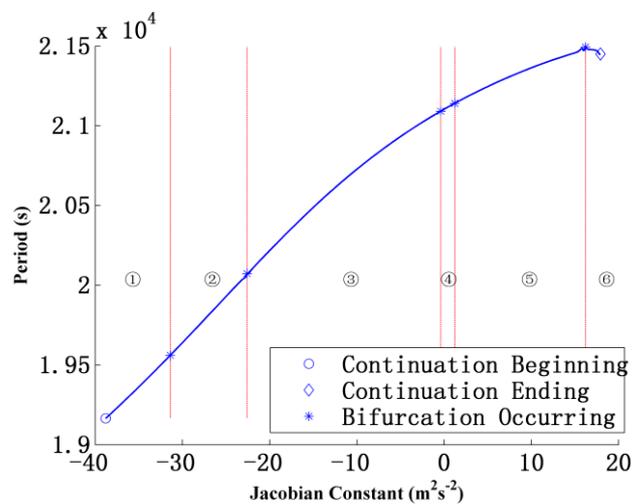

Figure 11. Illustration of the periodic orbit family continuation containing multiple bifurcations,

including the period's variation of orbits with a changing Jacobian constant and their relationship when bifurcations occurring. Topological cases in 6 zones are Case P1 Case P3, Case P4, Case P3, Case P4 and Case P2, respectively.

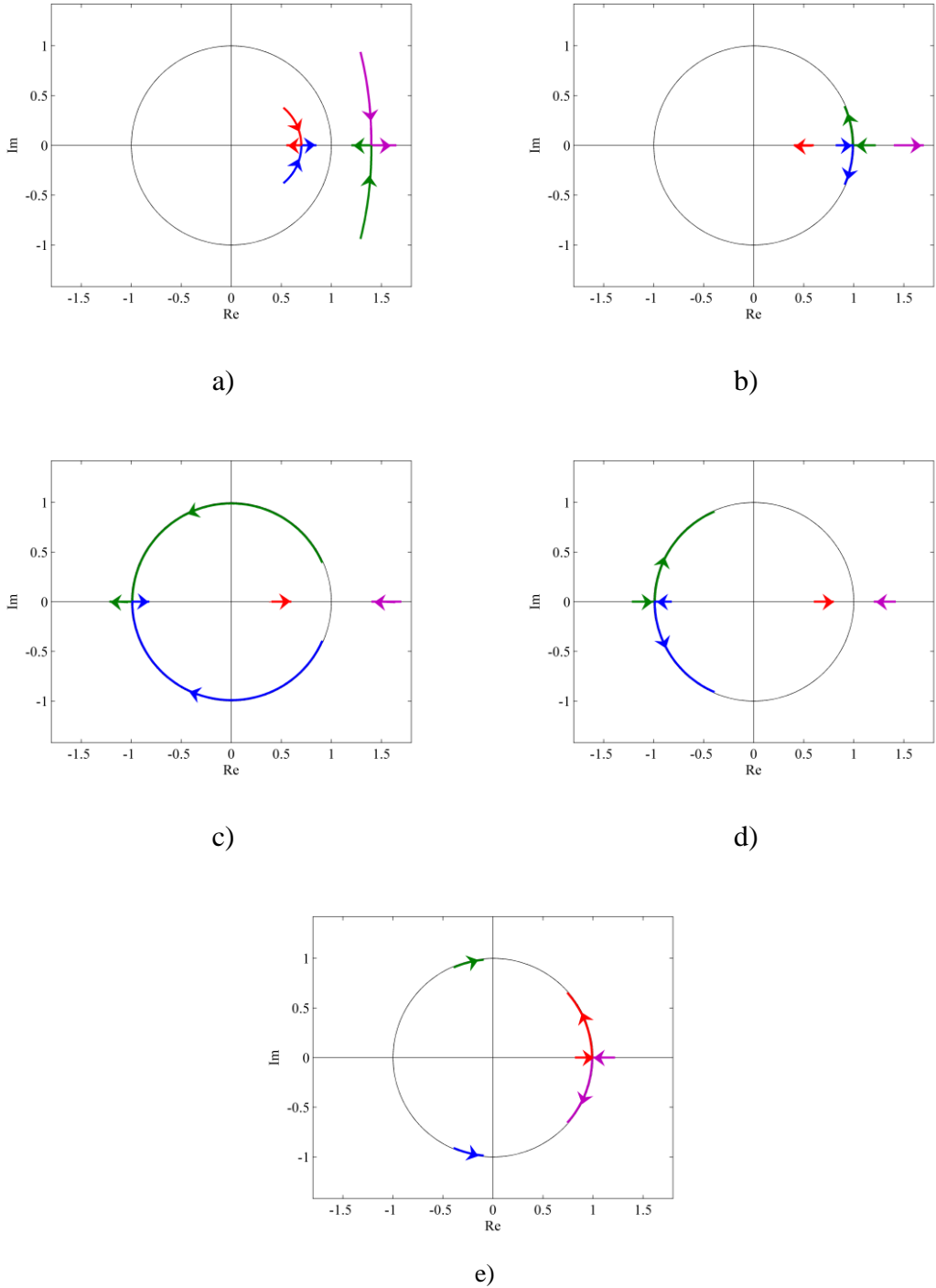

Figure 12. The topological transfer paths of the multiple bifurcations: a) shows the path of the real saddle bifurcation, b) shows the path of the first tangent bifurcation, c) shows the path of the first period-doubling bifurcation, d) shows the path of the second period-doubling bifurcation, and e) shows the path of the second tangent bifurcation.

During this continuation the Jacobian constants of the periodic orbits vary from $-38.79$ m$^2$s$^{-2}$ to $17.93$ m$^2$s$^{-2}$, and the periods of the periodic orbits vary from $1.9165\times10^4$ s to $2.1482\times10^4$ s. The real saddle bifurcation occurs when the Jacobian constant equals $-31.39$ m$^2$s$^{-2}$ and the period equals $1.9561\times10^4$ s. The first tangent bifurcation occurs when the Jacobian constant equals $-22.59$ m$^2$s$^{-2}$ and the period equals $2.0073\times10^4$ s. The first period-doubling bifurcation occurs when the Jacobian constant equals $-0.19$ m$^2$s$^{-2}$ and the period equals $2.1098\times10^4$ s. The second period-doubling bifurcation occurs when the Jacobian constant equals $1.01$ m$^2$s$^{-2}$ and the period equals $2.1135\times10^4$ s. The second tangent bifurcation occurs when the Jacobian constant equals $16.41$ m$^2$s$^{-2}$ and the period equals $2.1482\times10^4$ s. Figure (10a) shows the periodic orbits before the real saddle bifurcation; Figure (10b) shows the periodic orbits after the real saddle bifurcation and before the first tangent bifurcation; Figure (10c) shows the periodic orbits after the first tangent bifurcation and before the first period-doubling bifurcation; Figure (10d) shows the periodic orbits between the two period-doubling bifurcations; Figure (10e) shows the periodic orbits after the second period-doubling bifurcations and before the second tangent bifurcation; Figure (10f) shows the periodic orbits after the second tangent bifurcation. Figure 11 illustrates this continuation process using the relationship between Jacobian constant and the periodicity of the orbits.

Before the real saddle bifurcation the topological cases of the periodic orbits in the periodic orbit family belong to Case P1. During the continuation the topological case of the periodic orbit belongs to Case PDRS1 when the Jacobian constant equals −31.39 $m^2s^{-2}$ and the real saddle bifurcation occurs. After the bifurcation, the topological cases of the periodic orbits belong to Case P3. The periodic orbits remain unstable.

After the real saddle bifurcation the continuation goes on and the topological cases of the periodic orbits stay at Case P3 until the first tangent bifurcation occurs when the Jacobian constant equals −22.59 $m^2s^{-2}$. When the first tangent bifurcation appears, the topological case of the periodic orbit is Case P6. The topological cases of the periodic orbits after this bifurcation become Case P4, and these orbits are still unstable.

Between the first tangent bifurcation and the first period-doubling bifurcation, the topological cases of the periodic orbits in the periodic orbit family belong to Case P4. The topological case of the periodic orbit belongs to Case PPD4 when the Jacobian constant equals −0.19 $m^2s^{-2}$ and the first period-doubling bifurcation occurs. As mentioned before there is only one periodic orbit belonging to Case PPD4 during the bifurcation. After the bifurcation the topological cases of the periodic orbits revert back to Case P3. The periodic orbits remain unstable.

After that the topological cases of the periodic orbits won't change until the Jacobian constant reaches 1.01 $m^2s^{-2}$ and the second period-doubling bifurcation occurs. When this bifurcation appears, the topological case of the periodic orbit belongs to Case PPD4 again. The sojourn time at Case PPD4 is still zero as mentioned before, which indicates there is only one periodic orbit belonging to Case PPD4 during the bifurcation. After this the topological cases of the periodic orbits revert back to Case P4. The periodic orbits are still unstable.

Before the second tangent bifurcation, the periodic orbits' topological cases belong to Case P4. The continuation goes on and the second tangent bifurcation occurs when the Jacobian constant equals 16.41 $m^2s^{-2}$. When the second tangent bifurcation appears, the topological case of the periodic orbit is Case P5. The topological cases of the periodic orbits after this bifurcation belong to Case P2, and these periodic orbits become stable.

## 4. Conclusions

We discussed the periodic orbit families in the potential of an irregular body. Topological cases of periodic orbits and four kinds of basic bifurcations in periodic orbit families are analyzed in detail. Multiple bifurcations in periodic orbit families which consist of four kinds of basic bifurcations are investigated.

The binary period-doubling bifurcations and binary tangent bifurcations in periodic orbit families coexist in the same potential of the irregular asteroid 433 Eros. The binary period-doubling bifurcations in the periodic orbit family imply that the two stable regions for the motion around asteroid 433 Eros are separated by one unstable region. The trajectory of the periods and the Jacobian constants of the periodic orbit family with binary tangent bifurcations has a cusp when the first tangent bifurcation occurs, which implies that the trajectory is not smooth when the first tangent bifurcation occurs.

The triple bifurcations which consist of two real saddle bifurcations and one period-doubling bifurcation are also found; the characteristic multipliers of periodic orbits move from the positive half plane to the negative half plane between the first and the second real saddle bifurcations. The periodic orbit families continued from an equilibrium point of asteroid 433 Eros are found to have five bifurcations, i.e. one real saddle bifurcation, two tangent bifurcations, and two period-doubling bifurcations.

**Acknowledgements**

This research was supported by the National Science Foundation for Distinguished Young Scholars (11525208), the State Key Laboratory of Astronautic Dynamics Foundation (No. 2016ADL-0202), and the National Natural Science Foundation of China (No. 11372150).